\documentclass{article}

\usepackage{arxiv}
\usepackage[utf8]{inputenc}
\usepackage[T1]{fontenc}
\usepackage{hyperref}
\usepackage{url}
\usepackage{booktabs}
\usepackage{amsfonts}
\usepackage{amsmath}
\usepackage{amssymb}
\usepackage{nicefrac}
\usepackage{microtype}
\usepackage{graphicx}
\graphicspath{{./figures/}}
\usepackage[numbers,square,sort&compress]{natbib}
\usepackage{multirow}
\usepackage{float}
\usepackage[table,xcdraw]{xcolor}
\usepackage{enumitem}
\usepackage{array}
\usepackage{caption}
\usepackage{bm}

\title{Design Structure Matrix Modularization \\ with Large Language Models}

\date{A Preprint. Current Version: May 1, 2026.}

\author{
{
\hspace{1mm}Shuo Jiang}\thanks{Comments are welcome: \texttt{shuojiangcn@gmail.com}} \\
	Department of Systems Engineering \\
        City University of Hong Kong, Hong Kong \\
	\texttt{shuo.jiang@cityu.edu.hk} \\
    \And
	{\hspace{1mm}Jianxi Luo} \\
        Department of Systems Engineering \\
        City University of Hong Kong, Hong Kong \\
	\texttt{jianxi.luo@cityu.edu.hk} \\
}

\renewcommand{\headeright}{DSM Modularization with Large Language Models}
\renewcommand{\undertitle}{}
\renewcommand{\shorttitle}{}

\begin{document}
\maketitle

\begin{abstract}
Design Structure Matrix (DSM) modularization, the task of partitioning system elements into
cohesive modules, is a fundamental combinatorial challenge in engineering
design. Traditional methods treat modularization as a pure graph
optimization, without access to the engineering context embedded in the
system. Building on prior work on LLM-based combinatorial optimization
for DSM sequencing, this paper extends the method to modularization
across five cases and three backbone LLMs. Our method achieves
near-reference quality within 30 iterations without requiring specialized
optimization code. Counterintuitively, domain knowledge, beneficial in
sequencing, consistently impairs performance on more complex DSMs. We
attribute this to semantic misalignment between the LLM's functional
priors and the purely structural optimization objective, and propose the
\emph{semantic-alignment hypothesis} as a testable condition governing
knowledge effectiveness with LLMs. Ablation studies identify the most
effective input representation, objective formulation, and solution pool
design for practical deployment. These findings offer practical guidance
for deploying LLMs in engineering design optimization.
\end{abstract}


\section{Introduction}
\label{sec:intro}

Engineering systems of growing complexity require structured methods for
organizing and managing interdependencies among their constituent
elements. The Design Structure Matrix (DSM) has become a widely adopted
tool for this purpose, providing a compact matrix representation of
pairwise dependencies within a system \cite{Eppinger2012,Steward1981}.
DSM methods have been applied to product architecture, process planning,
project management, and organizational design
\cite{Langner2025,Luo2015,Roh2025,Solberg2025}. Many methodological literature
has developed around the core analytical tasks: sequencing,
partitioning, and modularization \cite{Browning2016}.

Among these tasks, DSM modularization (also referred to as clustering)
is particularly consequential for systems design. Modularization seeks
to partition the elements of a DSM into cohesive modules that minimize
coupling across module boundaries and maximize interaction density within
modules. Good modular architectures enable parallel development, support
reuse and upgrade, and reduce the propagation of design changes
\cite{Eppinger2016,Pimmler1994}. Thebeau formalized this
as an optimization problem via the TotalCost metric, which has been
adopted in many studies \cite{Thebeau2001}. The underlying combinatorial problem is
NP-hard: the number of feasible partitions grows as the Bell number
$B(n)$, reaching approximately $8\times10^{13}$ for a system with only
17 elements. This exponential growth motivates the development of
effective search heuristics.

Existing optimization algorithms, including Simulated Annealing, Genetic
Algorithms, and improved clustering heuristics, find near-optimal
partitions effectively
\cite{BorjessonHoltta2012,Thebeau2001,Yu2007}.  However, these methods
treat modularization as a pure graph optimization: they minimize the
objectives without reasoning about the engineering meaning of the
elements. As a result, they cannot incorporate designer knowledge about
functional relationships, system architecture intent, or physical
constraints. Recent advances in Large Language Models (LLMs) offer a
qualitatively different computational paradigm. Through in-context
learning and complex reasoning \cite{Brown2020}, LLMs can process both
structural information and natural language descriptions of the elements,
in principle combining optimization power with engineering reasoning.

Our prior work \cite{Jiang2025} demonstrated that LLM-based
combinatorial optimization (LLM-CO) can be effectively applied for DSM
sequencing, ordering activities to minimize feedback loops. We found
that LLM-CO is competitive with established benchmark methods, and
incorporating domain knowledge about activity/component semantics
significantly improved solution quality. DSM modularization presents a
distinct and more challenging setting for LLM-CO. The two tasks differ
fundamentally in their optimization objectives: sequencing minimizes
feedback loops, which map directly to workflow dependencies and are
semantically accessible to an LLM, while modularization optimizes a
structural graph metric with limited inherent semantic content. Whether
LLM-CO generalizes effectively to this setting, and whether domain
knowledge remains beneficial when semantic alignment between the LLM's
priors and the optimization objective cannot be assumed, are open
questions that motivate this study.

In this study, we apply LLM-CO to DSM modularization across five
engineering cases spanning aerospace, automotive, and mechanical
engineering domains, using three backbone LLMs. We address three
research questions:
\begin{itemize}
  \item[\textbf{RQ1}] Can LLM-CO achieve good modularization quality
    within a small iteration budget?
  \item[\textbf{RQ2}] How does domain knowledge affect performance, and
    why does it differ from its role in sequencing?
  \item[\textbf{RQ3}] Which input representations and design choices
    work best in practice with our LLM-CO framework?
\end{itemize}

The main contributions of this paper are:
\begin{enumerate}[label=(\arabic*)]
  \item We demonstrate LLM-CO as a viable approach for DSM
    modularization, achieving near-reference quality within 30 iterations
    across five engineering cases and three backbone LLMs, without
    requiring specialized optimization code.
  \item We identify a \emph{knowledge paradox}: domain knowledge,
    beneficial in DSM sequencing, systematically impairs modularization
    performance on more complex DSMs. We attribute this to semantic
    misalignment between the LLM's functional priors and the structural
    optimization objective, and propose this as a testable condition for
    knowledge effectiveness in LLM-CO more broadly.
  \item Through ablation studies on input representation, objective
    formulation, and solution pool design, we provide practical guidance
    for deploying LLM-CO in engineering design optimization.
\end{enumerate}

The remainder of this paper is organized as follows. Section~\ref{sec:bg}
reviews background on DSM modularization and LLMs for combinatorial
optimization. Section~\ref{sec:method} presents the LLM-CO framework.
Section~\ref{sec:exp} describes the experimental setup.
Section~\ref{sec:results} presents experimental results and discusses the findings.
Section~\ref{sec:conclusion} concludes the paper.

\section{Background}
\label{sec:bg}

\subsection{DSM Modularization}

The Design Structure Matrix was introduced by \cite{Steward1981} as a
square matrix representation of directed dependencies among system
elements. A non-zero entry $\text{DSM}[i][j]$ indicates that element
$i$ receives input from element $j$, with the entry value representing
the strength of that dependency. DSMs come in three principal types
depending on what the rows and columns represent: activity-based DSMs
encode information flows between design activities; parameter-based DSMs
represent functional dependencies among design parameters; and
component-based DSMs capture physical interfaces between hardware
components \cite{Browning2001,Eppinger2012}. The present study includes
all three types.

Pimmler and Eppinger \cite{Pimmler1994} established a framework for quantifying interactions
across spatial, energy, information, and material flows, providing the
theoretical basis for assigning numerical interaction strengths in a DSM.
Thebeau \cite{Thebeau2001} built on this to formalize modularization as an
optimization problem, introducing TotalCost as the objective function and
proposing a stochastic hill-climbing algorithm to optimize this objective. Later
work has explored broader algorithmic approaches. For example,
Yu et al.\ \cite{Yu2007} applied genetic algorithms to exploit global search
capabilities. Börjesson and Hölttä-Otto \cite{BorjessonHoltta2012} developed an improved
clustering algorithm that avoids some local optima of the original
hill-climbing method. These algorithms have been validated on a range of
industrial DSMs and represent the current state of practice. However,
such methods treat modularization as a pure graph optimization without
incorporating engineering knowledge.

\subsection{LLMs for Combinatorial Optimization}

Combinatorial optimization has a long history in operations research and
computer science, ranging from exact methods (e.g., branch-and-bound,
dynamic programming) to metaheuristics (e.g., simulated annealing,
evolutionary algorithms). Recent work has explored a new paradigm: using
LLMs as optimizers. Yang et al.\ \cite{Yang2024} introduced OPRO, demonstrating that
LLMs can optimize objective functions by iteratively refining candidate
solutions described in natural language. Given a prompt containing
previously evaluated solutions ranked by quality, the LLM generates
improved candidates by extrapolating patterns from the examples. This
approach requires no gradient computation or explicit solver code,
relying instead on the LLM's pattern-completion capabilities. FunSearch
\cite{RomeraParedes2024} extended this idea to mathematical discovery,
using LLMs to search over program space for functions that improve on
known solutions to hard combinatorial problems. Their results demonstrated
that LLM-guided search can discover state-of-the-art solutions on problems
including cap set construction and bin packing. These approaches share a
common structure: an LLM generates candidate solutions, which are
evaluated by an external scorer, with the best solutions fed back as
in-context examples for the next generation round.

In the DSM domain, two recent lines of work apply LLMs to DSM tasks.
Koh \cite{Koh2024} introduced Auto-DSM, demonstrating that LLMs can
generate DSM structures from natural language project descriptions, a
generation task rather than optimization. A follow-up study
\cite{Koh2026} further evaluated LLM-based extraction of indirect
design dependencies, examining how writing style and entity naming affect
retrieval accuracy across five LLMs. Together, these studies establish
that LLMs can interpret and produce DSM content from text, though they
do not address modularization. In our previous study \cite{Jiang2025},
we applied LLM to DSM sequencing, finding that LLM-CO is competitive with
established methods and that domain knowledge about activity semantics
consistently improved solution quality. The present work extends these
findings to DSM modularization, where the objective function differs
fundamentally in its semantic interpretability.

\section{LLM-CO Framework for DSM Modularization}
\label{sec:method}

\subsection{Problem Formulation}

Given a DSM with $n$ elements represented as a weighted directed graph,
the modularization problem seeks a partition $\mathcal{M} =
\{M_1,\ldots,M_K\}$ of elements into $K$ modules ($2 \leq K \leq n$) that
minimizes a structural cost objective (defined formally in
Section~\ref{subsec:metrics}). Because the global optimum is
computationally intractable for large $n$, we establish a reference
benchmark by running Simulated Annealing (SA) with 10{,}000 independent
random restarts and taking the best solution found; we refer to this as
the \emph{SA reference} (computational details in Appendix~\ref{app:sa}).
The SA reference is not provably optimal but represents the best-known
solution under this metric and provides a stable, reproducible reference
for comparison.

\subsection{LLM-CO Framework}

\begin{figure}
  \centering
  \includegraphics[width=\linewidth]{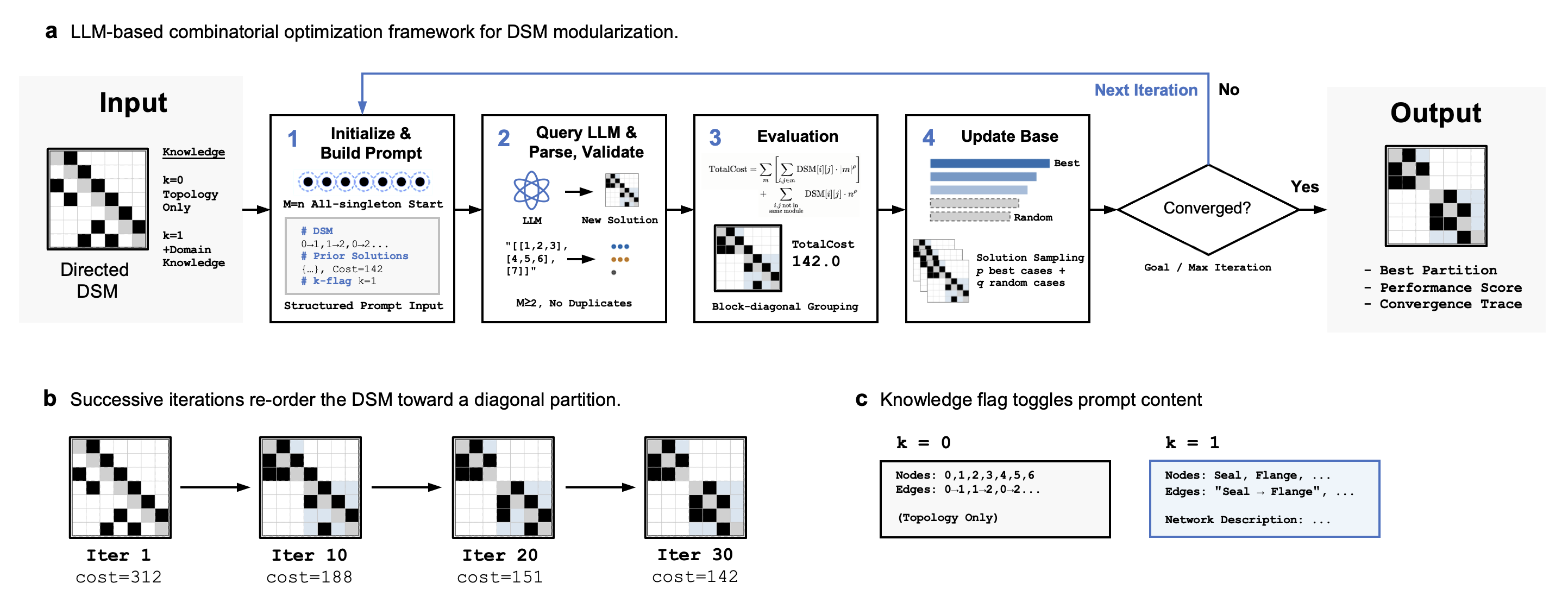}
  \caption{LLM-based combinatorial optimization framework for DSM
  modularization. (a)~The iterative loop consists of four stages: prompt
  construction, LLM query and validation, evaluation, and solution pool
  update. (b)~Successive iterations re-order the DSM toward a
  block-diagonal partition. (c)~Knowledge flag $k$ controls prompt
  content.}
  \label{fig:framework}
\end{figure}

Figure~\ref{fig:framework} illustrates the proposed LLM-CO framework.
The iterative procedure consists of four stages: \textbf{(1) Initialize and build prompt.} At the first iteration,
    each element is assigned to its own singleton module, yielding $n$
    modules of size 1. The current solution pool is then encoded into a
    structured prompt containing the DSM representation, the ranked prior
    solutions, and optionally domain knowledge about the elements.
\textbf{(2) Query LLM and validate.} The LLM is queried to generate
    a new candidate partition, expressed as a mapping from element
    identifiers to module labels; the response is checked for structural
    validity, requiring all elements to be assigned to exactly one
    module with at least two modules total.
\textbf{(3) Evaluation.} The cost objective is computed for the
    validated partition.
\textbf{(4) Update solution pool.} The pool retains the top-ranked
    solutions found so far, supplemented by randomly sampled solutions
    from the history to maintain diversity; the best-ever solution is
    tracked separately as the final output.
This loop repeats for a fixed number of iterations, with the best
solution found across all iterations returned as the final output.

Each prompt is structured around three components. The first is the
\textbf{DSM representation}. The DSM is encoded as a directed edge list:
a set of $(i,\,j,\,\text{weight})$ triples indicating that element $i$
receives input from element $j$ with a given strength. Node identifiers
are shuffled randomly at each iteration to prevent position bias. This
format was selected based on ablation results (Section~\ref{subsec:ablation}),
as it provides complete directional information while remaining compact
and reliably parseable by all three LLMs tested. The second component is
the \textbf{solution base}. Previously found partitions are presented in
ranked order from best to worst, providing the LLM with concrete examples
from which to infer the optimization direction. In our default
configuration, the pool contains $p=5$ best solutions found so far and
$q=5$ randomly sampled solutions, balancing exploitation of high-quality
regions and exploration of the broader solution space. Each solution is
presented as a module-assignment list together with its objective value.
The third component is a knowledge flag that determines whether
engineering semantic content is included in the prompt. Under the
knowledge-off condition ($k=0$), node identifiers are replaced with
anonymized labels (e.g., N1, N2, \ldots) and no natural language
descriptions are provided, so the LLM receives only topological
information. Under the knowledge-on condition ($k=1$),
each node is identified by its engineering name and accompanied by a
brief functional description. The comparison between $k=0$ and $k=1$
forms the central experimental contrast of this study. Prompt details
are provided in Appendix~\ref{app:prompt}.

\section{Experimental Setup}
\label{sec:exp}

\subsection{DSM Cases}

We evaluate LLM-CO on five engineering DSM cases, summarized in
Table~\ref{tab:cases}, where $N$ denotes the number of nodes, $E$ the
number of directed edges, and Density $= E/(N \times (N-1))$. The cases
range from 12 to 19 elements, spanning activity-based, parameter-based,
and component-based DSM types
across aerospace, automotive, and mechanical engineering domains. All
five cases were collected from peer-reviewed literature. For each case,
three types of information were extracted from the original reference:
the names of all system elements, the edge list derived from the
adjacency matrix, and a natural language description of each element's
function within the system. These cases extend those used in our prior
study on LLM-CO for DSM sequencing \cite{Jiang2025}.

\begin{table}[H]
\caption{Five Benchmark DSM Cases}
\label{tab:cases}
\centering
\begin{tabular}{lcccccl}
\toprule
\textbf{Case} & $\boldsymbol{N}$ & $\boldsymbol{E}$ &
\textbf{Density} & \textbf{DSM Type} & \textbf{Domain} &
\textbf{Source} \\
\midrule
UCAV              & 12 & 52  & 0.40 & Activity  & Aerospace       & \cite{Browning1998} \\
Kodak Cartridge   & 13 & 49  & 0.32 & Parameter & Automotive      & \cite{Eppinger2016} \\
Brake System      & 14 & 35  & 0.19 & Activity  & Automotive      & \cite{Black1990}    \\
HeatEx            & 17 & 43  & 0.16 & Parameter & Mechanical Eng. & \cite{Amen1999}     \\
Helicopter        & 19 & 109 & 0.32 & Component & Aerospace       & \cite{Clarkson2004} \\
\bottomrule
\end{tabular}
\end{table}

\subsection{Experimental Settings}

We evaluate three backbone LLMs: Claude-Sonnet-4.6 (Hereafter,
\textbf{Claude}) \cite{Anthropic2026}, GPT-5.2 (Hereafter,
\textbf{GPT}) \cite{OpenAI2025}, and Qwen-3.5-Plus (Hereafter,
\textbf{Qwen}) \cite{QwenTeam2026}. These models represent distinct
architectures and training regimes, allowing us to assess the robustness
of findings across model families. For each combination of model, DSM
case, and knowledge condition ($k=0$ and $k=1$), we run 10 independent
trials of 30 iterations each. All experiments use the default solution
pool configuration ($p=5$ best, $q=5$ random) and directed edge list
input format unless otherwise specified. Temperature is set to 1.0 for
all models to encourage solution diversity. More implementation
details are provided in Appendix~\ref{app:hyperparams}.

\subsection{Evaluation Metrics}
\label{subsec:metrics}

We use two metrics to evaluate modularization quality.
\textbf{TotalCost} \cite{Thebeau2001} is the primary optimization
objective:
\begin{equation}
  \text{TotalCost} \;=\;
  \sum_{k} \left(\sum_{\substack{i,j \in M_k}} w_{ij}\right)
  \frac{|M_k|^\rho}{n^\rho}
  \;+\; n \sum_{\substack{i \in M_a,\; j \in M_b \\ a \neq b}} w_{ij}
  \label{eq:totalcost}
\end{equation}
where $w_{ij}$ is the interaction weight between elements $i$ and $j$,
$|M_k|$ is the size of module $M_k$, $n$ is the total number of
elements, and $\rho$ is a size-penalty exponent (set to $\rho=1$ throughout).
Lower TotalCost is better: the formulation rewards dense intra-module
interactions (multiplied by $|M_k| < n$) and penalizes cross-module
interactions.

\textbf{Clustering Efficiency (CE)} is a secondary metric defined as
the fraction of total weighted interactions that fall within modules:
\begin{equation}
  \text{CE} \;=\;
  \frac{\displaystyle\sum_{\substack{i,j \in \text{same module}}}
  w_{ij}}{\displaystyle\sum_{i,j} w_{ij}}
  \label{eq:ce}
\end{equation}
CE ranges from 0 to 1, and higher values indicate better modularization.
CE is reported alongside TotalCost in Table~\ref{tab:results} as a
complementary measure of solution quality.

Unless otherwise noted, all metrics are aggregated as mean~$\pm$
standard deviation across 10 independent runs. For convergence analysis,
we additionally report
\begin{equation}
  \text{Gap\%} \;=\; \frac{\text{TotalCost} - \text{SA reference}}
  {\text{SA reference}} \times 100\%
  \label{eq:gap}
\end{equation}
where a Gap\% of $0\%$ indicates that the run matched the SA reference.

\subsection{Ablation Conditions}
\label{subsec:ablation-design}

To understand what shapes LLM-CO performance across key design choices,
we conduct ablation studies on three dimensions, all using the UCAV case
with the Claude model. Domain knowledge is included ($k=1$) in all
ablation conditions. This case and model were selected because they
represent a mid-complexity setting where all three ablation dimensions
produce observable performance differences.

For \textit{solution pool design}, we compare three variants: the
balanced baseline ($p=5$ best, $q=5$ random), an exploitation-only
variant ($p=5$ best, $q=0$), and an exploration-only variant ($p=0$,
$q=5$ random). For \textit{objective formulation}, we compare the
baseline prompt against a simple prompt with the explicit TotalCost
formula removed, retaining only the natural language task description
and ranked solution examples. For \textit{input representation}, we
compare the directed edge list baseline against an adjacency matrix, an
undirected edge list (directionality removed), and a natural language
description of dependencies.

\section{Results and Discussion}
\label{sec:results}

\subsection{Modularization Quality and Convergence Behavior}

Table~\ref{tab:results} summarizes mean TotalCost and Clustering
Efficiency (CE) across 10 independent runs for all three models and
knowledge conditions. Claude under $k=0$ achieves the most consistent
performance: on the largest case (Helicopter, $n=19$), all 10 runs
match the SA reference in TotalCost ($1371\pm0$), while CE
($0.675\pm0.011$) exceeds the SA reference value of 0.514, indicating
that LLM-CO identifies equally optimal partitions with higher
intra-module interaction density. Notably, incorporating domain
knowledge does not consistently improve performance here, in contrast to
our earlier finding in DSM sequencing where knowledge was uniformly
beneficial \cite{Jiang2025}. Across the three intermediate cases
(Kodak, Brake, Heat Exchanger), Claude $k=0$ remains within 1.3\% of the SA
reference. GPT and Qwen achieve competitive results on smaller cases but
exhibit substantially higher run-to-run variance, particularly on UCAV
and Helicopter. Across all three models, the results confirm that LLM-CO
is a viable approach for DSM modularization without writing any solver
code.

\begin{table}[H]
\caption{Performance Across Models, Knowledge Conditions, and DSM Cases.
  Values show TotalCost mean~$\pm$ std (CE mean~$\pm$ std) over 10 runs.
  \textbf{with} = $k=1$; \textbf{without} = $k=0$.}
\label{tab:results}
\centering
\setlength{\tabcolsep}{4pt}
\renewcommand{\arraystretch}{1.2}
\begin{tabular}{llccccc}
\toprule
\textbf{Model} & \textbf{Know.} &
  \textbf{UCAV} & \textbf{Kodak} & \textbf{Brake} &
  \textbf{HeatEx} & \textbf{Helicopter} \\
\midrule
\multirow{2}{*}{Claude-Sonnet-4.6}
  & with
    & $\bm{450.0\pm0.0}$ & $437.8\pm0.4$ & $292.4\pm3.7$
    & $433.8\pm7.4$ & $1420.8\pm36.6$ \\
  & & $(0.56\pm0.00)$ & $(0.64\pm0.01)$ & $(0.63\pm0.00)$
    & $(0.70\pm0.07)$ & $(0.59\pm0.10)$ \\[2pt]
  & without
    & $454.8\pm10.5$ & $\bm{436.7\pm2.2}$ & $291.6\pm3.2$
    & $\bm{412.0\pm9.1}$ & $\bm{1371.0\pm0.0}$ \\
  & & $(0.53\pm0.05)$ & $(0.59\pm0.07)$ & $(0.63\pm0.00)$
    & $(0.73\pm0.04)$ & $(0.68\pm0.01)$ \\
\midrule
\multirow{2}{*}{GPT-5.2}
  & with
    & $517.6\pm53.3$ & $462.2\pm25.1$ & $300.4\pm13.4$
    & $448.0\pm0.0$ & $1391.9\pm30.0$ \\
  & & $(0.72\pm0.17)$ & $(0.74\pm0.11)$ & $(0.66\pm0.07)$
    & $(0.65\pm0.00)$ & $(0.64\pm0.06)$ \\[2pt]
  & without
    & $503.1\pm42.4$ & $461.3\pm10.1$ & $\bm{290.0\pm0.0}$
    & $457.9\pm24.5$ & $1481.1\pm176.7$ \\
  & & $(0.68\pm0.10)$ & $(0.79\pm0.10)$ & $(0.63\pm0.00)$
    & $(0.77\pm0.09)$ & $(0.70\pm0.08)$ \\
\midrule
\multirow{2}{*}{Qwen-3.5-Plus}
  & with
    & $538.9\pm56.7$ & $444.8\pm8.3$ & $298.3\pm0.5$
    & $441.8\pm11.2$ & $1543.8\pm44.0$ \\
  & & $(0.81\pm0.17)$ & $(0.66\pm0.03)$ & $(0.62\pm0.01)$
    & $(0.64\pm0.03)$ & $(0.41\pm0.05)$ \\[2pt]
  & without
    & $526.0\pm38.7$ & $442.4\pm6.7$ & $298.8\pm15.8$
    & $483.0\pm58.4$ & $1406.9\pm42.6$ \\
  & & $(0.72\pm0.15)$ & $(0.53\pm0.05)$ & $(0.66\pm0.07)$
    & $(0.83\pm0.07)$ & $(0.65\pm0.06)$ \\
\midrule
SA reference (10K Best) & -- & $446\ (0.44)$ & $434\ (0.51)$ & $290\ (0.63)$
               & $407\ (0.74)$ & $1371\ (0.51)$ \\
\bottomrule
\end{tabular}
\end{table}

\begin{figure}
  \centering
  \includegraphics[width=\linewidth]{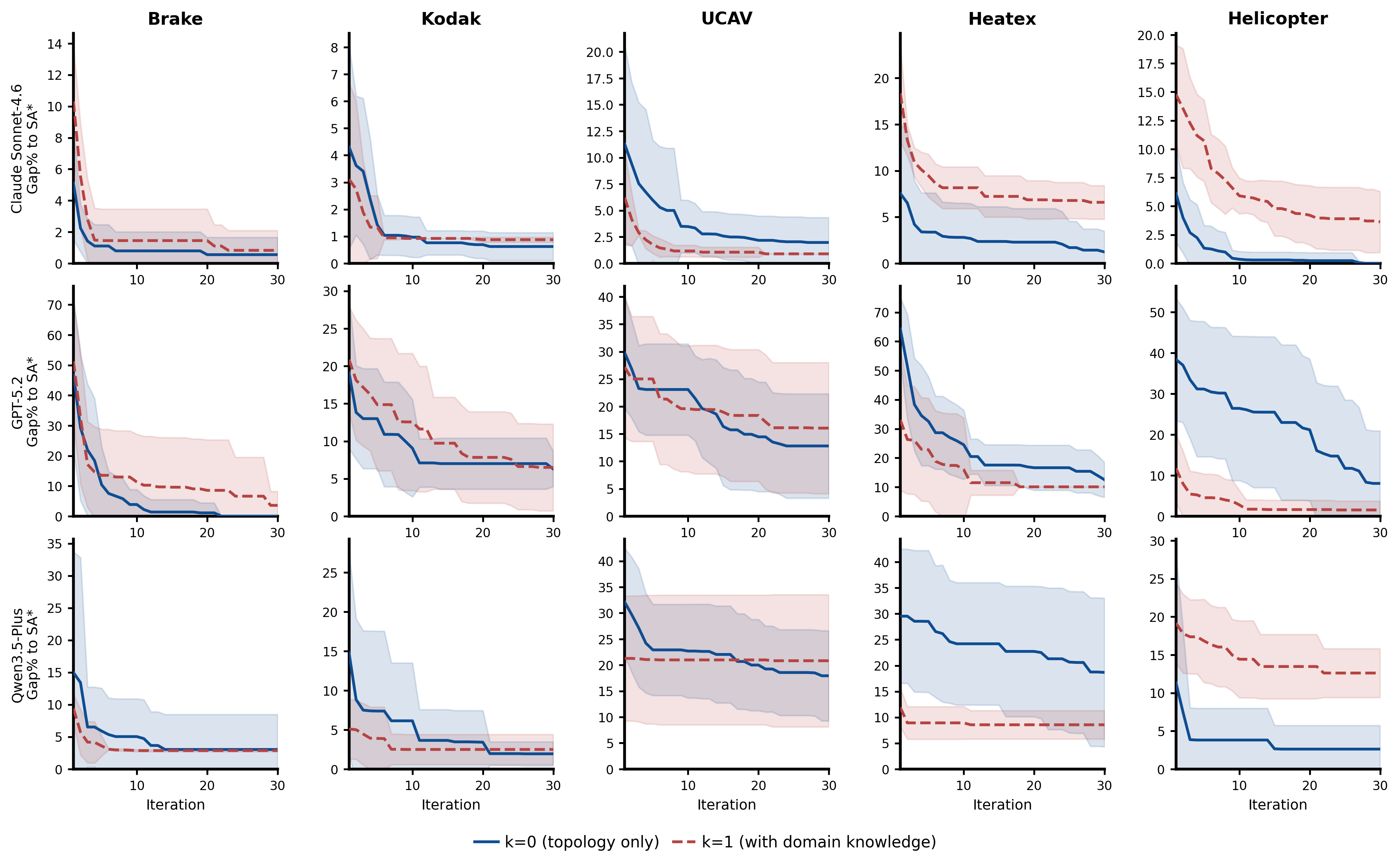}
  \caption{Convergence behavior of LLM-CO across three LLMs and five
  DSM cases. Each panel shows Gap\% (mean $\pm$ std over 10 runs) vs.\
  iteration number, for $k=0$ and $k=1$ conditions.}
  \label{fig:convergence}
\end{figure}

Figure~\ref{fig:convergence} shows the convergence behavior of all three
models across all five cases. Claude converges rapidly and stably:
Gap\% drops sharply within the first five iterations and approaches a
plateau well before iteration 30, with narrow standard deviation bands
across all cases. GPT and Qwen present a contrasting picture: convergence
is slower, bands are substantially wider, and several panels show curves
still declining at iteration 30, particularly on larger cases. This
suggests that while a 30-iteration budget is sufficient for Claude, GPT
and Qwen may benefit from additional iterations on more complex DSMs.
The difference in convergence stability across models is consistent with
the variance patterns reported in Table~\ref{tab:results}, and reinforces
Claude as the most reliable backbone for DSM modularization within the
tested iteration budget.

\subsection{The Knowledge Paradox: When Domain Information Hurts}

\begin{figure}
  \centering
  \includegraphics[width=\linewidth]{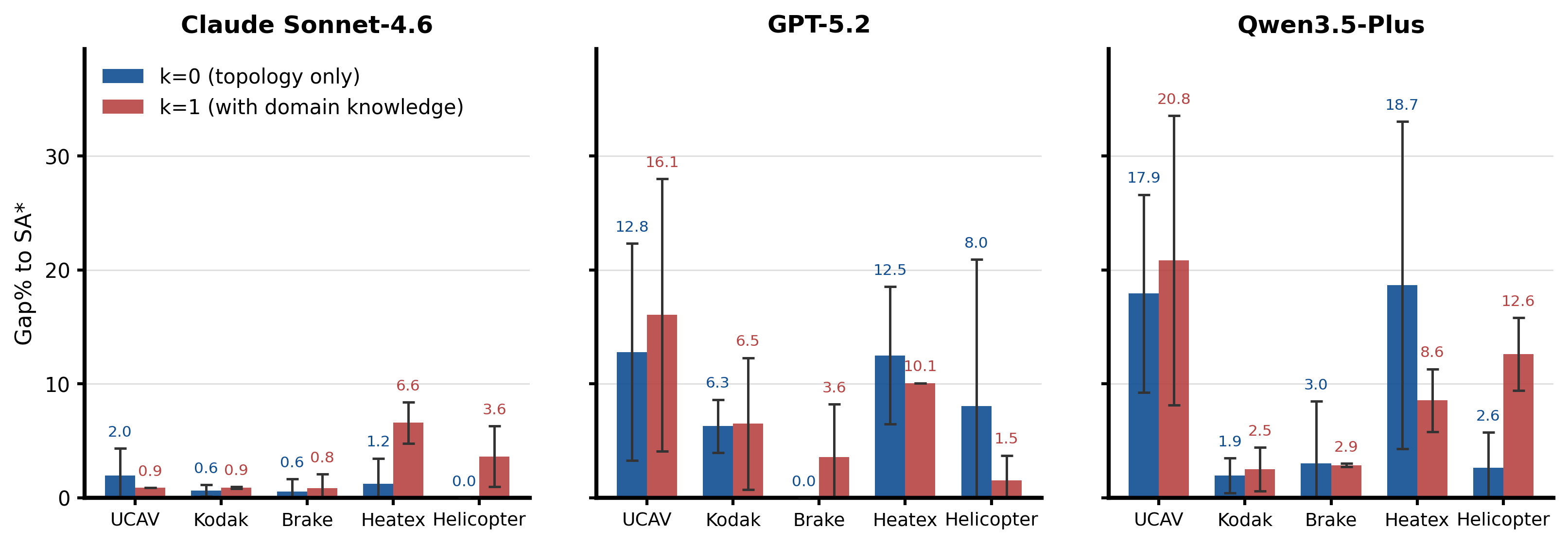}
  \caption{Modularization quality with and without domain knowledge.
  Each bar shows final Gap\% for $k=1$ (with) and $k=0$ (without)
  across models and cases.}
  \label{fig:knowledge}
\end{figure}

The effect of domain knowledge on performance depends strongly on both
DSM complexity and model capability, as shown jointly by the final Gap\%
results (Figure~\ref{fig:knowledge}) and the convergence trajectories
(Figure~\ref{fig:convergence}). For Claude, the pattern is clear: domain
knowledge shows mixed or negligible effects on the three simpler cases,
with a marginal improvement on UCAV (Gap\% of 0.9\% with vs.\ 2.0\%
without domain knowledge) and essentially no change on Kodak and Brake.
On the two larger cases, however, domain knowledge consistently and
substantially impairs performance, with HeatEx incurring a 5.4\%
penalty and Helicopter a 3.6\% penalty. On the Helicopter case, Claude
$k=0$ achieves 10 out of 10 SA-reference-matching runs while $k=1$
yields a mean Gap\% of 3.6\% with zero runs reaching the reference. The
convergence curves in Figure~\ref{fig:convergence} reinforce this
finding: for Claude, the separation between $k=0$ and $k=1$ is
established within the first few iterations and maintained throughout,
indicating that knowledge shapes the search trajectory from the outset
rather than imposing a transient initial penalty.

For GPT and Qwen, the convergence trajectories in Figure~\ref{fig:convergence} reveal a more nuanced
pattern: $k=1$ tends to reach lower Gap\% in early iterations,
suggesting that domain knowledge provides a useful shortcut when the
model cannot yet identify high-quality structural partitions on its own.
However, $k=0$ catches up or surpasses $k=1$ in later iterations across
most cases, indicating that semantic priors ultimately constrain rather
than guide the search. This interaction between knowledge and model
capability supports a capability-moderation interpretation: for stronger
models like Claude, structural optimization under $k=0$ is reliable
enough that semantic priors serve only as a distraction; for weaker
models, knowledge offers an early advantage that diminishes as the
solution pool accumulates informative examples.

We attribute the impairment to \emph{semantic misalignment} between the
LLM's domain priors and the optimization objective. When node names are
visible, the LLM applies engineering intuitions about functional grouping
that are intuitively reasonable but diverge from the structural optimum.
For simpler DSMs, functional and structural groupings tend to coincide
and the effect is negligible; for more complex DSMs, the two diverge and
the penalty becomes consistent. This stands in contrast to DSM sequencing
\cite{Jiang2025}, where minimizing feedback loops aligns directly with
engineering process logic and domain knowledge consistently helps.

\subsection{Ablation Analysis}
\label{subsec:ablation}

\begin{figure}[H]
  \centering
  \includegraphics[width=\linewidth]{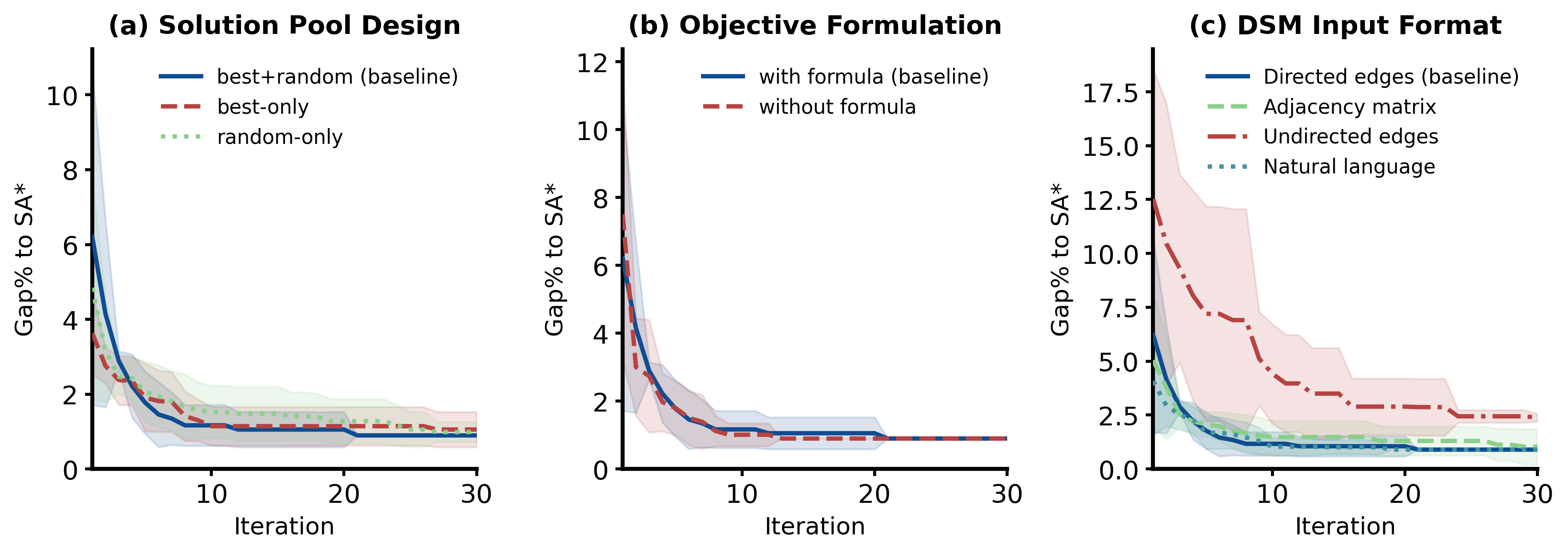}
  \caption{Ablation convergence behavior (UCAV, Claude, $k=1$).
  (a)~Solution pool design. (b)~Objective formulation.
  (c)~DSM input format.}
  \label{fig:ablation}
\end{figure}

Figure~\ref{fig:ablation} examines three design dimensions of the
LLM-CO framework, all using UCAV with Claude under $k=1$. The
solution pool design has a clear effect on both final quality
and convergence stability (Figure~\ref{fig:ablation}a). The balanced
pool (5 best + 5 random) achieves the lowest final Gap\% with the
narrowest standard deviation band, outperforming both the
exploitation-only (best-only) and exploration-only (random-only)
variants. The exploitation-only variant converges quickly in early
iterations but with higher run-to-run variance, while the
exploration-only variant shows slower early improvement in the absence
of high-quality exemplars. These results suggest that maintaining both
exploitation of high-quality solutions and exploration of the broader
solution space is important for stable and effective LLM-CO.

Removing the explicit objective formula from the prompt has no
measurable effect: the two conditions produce near-identical convergence
curves throughout all 30 iterations (Figure~\ref{fig:ablation}b). This
indicates that the LLM does not use the formula for numerical evaluation
but instead infers the optimization direction from the ranked solution
examples alone, consistent with in-context learning as implicit
optimization \cite{Yang2024}. A practical consequence is that LLM-CO
can be applied to problems where the objective is difficult to specify
analytically, provided that candidate solutions can be evaluated and
ranked externally.

Among the input representation variants
(Figure~\ref{fig:ablation}c), the directed edge list, adjacency matrix,
and natural language description all converge to similarly low Gap\% by
iteration 30, but the undirected edge list remains clearly worse
throughout. Removing directionality from the edge representation is the
single most damaging design choice tested, demonstrating that dependency
direction is a critical signal for LLM-CO: knowing which element depends
on which, rather than merely that two elements are connected,
substantially improves solution quality. The directed edge list is
recommended as the default format for its combination of directional
completeness, compactness, and parsing reliability.

\section{Concluding Remarks}
\label{sec:conclusion}

This paper presents a systematic investigation of LLM-based
combinatorial optimization for DSM modularization across five
engineering cases and three backbone LLMs. LLM-CO achieves near-reference
quality within 30 iterations without solver code or gradient computation,
with Claude-Sonnet-4.6 demonstrating the most consistent performance
across all cases. The central finding is a \emph{knowledge paradox}:
domain knowledge, beneficial in DSM sequencing with LLMs, systematically
impairs modularization performance on more complex DSMs. We attribute
this to semantic misalignment between the LLM's functional priors and
the purely structural optimization objective. Across models, we find
that this paradox is most clearly observable in stronger models capable
of reliable structural optimization, while less capable models show an
early advantage from domain knowledge that diminishes as the solution
pool matures.

This study has several limitations. The benchmark cases cover DSMs with
at most 19 elements, and scalability to larger instances remains an open
question. The SA reference is not provably optimal for the larger cases,
and additional model families beyond the three tested may exhibit
different knowledge sensitivity patterns. Different DSM types may exhibit
different patterns of semantic alignment, and the knowledge effect may
vary accordingly. For instance, Sosa et al.\ \cite{Sosa2004} demonstrated empirically
that misalignment between component architecture and organizational
structure can be substantial and difficult to detect, suggesting that
domain knowledge about physical interfaces may be more tacit and harder
for LLMs to leverage effectively than knowledge about activities or
parameters. Systematically investigating this across DSM types requires
broader empirical coverage than the present study provides. Finally, this
study evaluates modularization quality using TotalCost and Clustering
Efficiency, which are commonly used in the engineering design community;
future work should examine whether findings generalize to alternative
metrics from the network science community, such as the modularity metric
described in \cite{LeichtNewman2008}.

Together, the findings presented in this paper establish LLM-CO as a viable and
flexible approach for engineering design optimization, one that requires
no gradient computation or solver code and can be adapted to new
objectives by updating the evaluation function alone. More broadly, the
semantic-alignment hypothesis offers a testable principle for
anticipating when domain knowledge will help or hurt in LLM-based
optimization, with implications extending beyond DSMs to a wider class
of combinatorial problems in engineering design.

\bibliographystyle{unsrtnat}
\bibliography{references}

@inproceedings{Amen1999,
  author    = {Rafael Amen and Ingvar Rask and Staffan Sunnersjö},
  title     = {Matching Design Tasks to Knowledge-Based Software Tools: When Intuition Does Not Suffice},
  booktitle = {Proceedings of the ASME International Design Engineering Technical Conferences and Computers and Information in Engineering Conference (IDETC/CIE 1999)},
  pages     = {1165--1174},
  year      = {1999},
  publisher = {ASME}
}

@misc{Anthropic2026,
  author = {Anthropic},
  title  = {{Claude Sonnet 4.6 System Card}},
  year   = {2026},
  howpublished = {\url{https://www.anthropic.com/news/claude-sonnet-4-6}}
}

@techreport{Black1990,
  author      = {Thomas A. Black and Charles H. Fine and Emanuel M. Sachs},
  title       = {A Method for Systems Design Using Precedence Relationships: An Application to Automotive Brake Systems},
  institution = {Sloan School of Management, Massachusetts Institute of Technology},
  year        = {1990},
  type        = {Working Paper}
}

@inproceedings{BorjessonHoltta2012,
  author    = {Fredrik Börjesson and Katja Hölttä-Otto},
  title     = {Improved Clustering Algorithm for Design Structure Matrix},
  booktitle = {Proceedings of the ASME 2012 International Design Engineering Technical Conferences \& Computers and Information in Engineering Conference (IDETC/CIE 2012)},
  pages     = {921--930},
  year      = {2012},
  doi       = {10.1115/DETC2012-70076}
}

@inproceedings{Brown2020,
  author    = {Tom B. Brown and Benjamin Mann and Nick Ryder and Melanie Subbiah and Jared D. Kaplan and Prafulla Dhariwal and Arvind Neelakantan and Pranav Shyam and Girish Sastry and Amanda Askell and Sandhini Agarwal and Ariel Herbert-Voss and Gretchen Krueger and Tom Henighan and Rewon Child and Aditya Ramesh and Daniel M. Ziegler and Jeffrey Wu and Clemens Winter and Christopher Hesse and Mark Chen and Eric Sigler and Mateusz Litwin and Scott Gray and Benjamin Chess and Jack Clark and Christopher Berner and Sam McCandlish and Alec Radford and Ilya Sutskever and Dario Amodei},
  title     = {Language Models are Few-Shot Learners},
  booktitle = {Advances in Neural Information Processing Systems},
  pages     = {1877--1901},
  year      = {2020},
  publisher = {Curran Associates, Inc.}
}

@article{Browning2016,
  author  = {Tyson R. Browning},
  title   = {Design Structure Matrix Extensions and Innovations: A Survey and New Opportunities},
  journal = {IEEE Transactions on Engineering Management},
  volume  = {63},
  pages   = {27--52},
  year    = {2016},
  doi     = {10.1109/TEM.2015.2491283}
}

@article{Browning2001,
  author  = {Tyson R. Browning},
  title   = {Applying the Design Structure Matrix to System Decomposition and Integration Problems: A Review and New Directions},
  journal = {IEEE Transactions on Engineering Management},
  volume  = {48},
  pages   = {292--306},
  year    = {2001},
  doi     = {10.1109/17.946528}
}

@phdthesis{Browning1998,
  author = {Tyson R. Browning},
  title  = {Modeling and Analyzing Cost, Schedule, and Performance in Complex System Product Development},
  school = {Sloan School of Management, Massachusetts Institute of Technology},
  year   = {1998}
}

@article{Clarkson2004,
  author  = {P. John Clarkson and Caroline Simons and Claudia Eckert},
  title   = {Predicting Change Propagation in Complex Design},
  journal = {Journal of Mechanical Design},
  volume  = {126},
  pages   = {788--797},
  year    = {2004},
  doi     = {10.1115/1.1765117}
}

@book{Eppinger2012,
  author    = {Steven D. Eppinger and Tyson R. Browning},
  title     = {Design Structure Matrix Methods and Applications},
  publisher = {MIT Press},
  address   = {Cambridge, MA},
  year      = {2012}
}

@book{Eppinger2016,
  author    = {Steven D. Eppinger and Karl T. Ulrich},
  title     = {Product Design and Development},
  edition   = {6th},
  publisher = {McGraw-Hill Education},
  address   = {New York, NY},
  year      = {2016}
}

@article{Jiang2025,
  author  = {Shuo Jiang and Min Xie and Jianxi Luo},
  title   = {Large Language Models for Combinatorial Optimization of Design Structure Matrix},
  journal = {Proceedings of the Design Society},
  volume  = {5},
  year    = {2025},
  doi     = {10.1017/pds.2025.10234}
}

@article{Koh2026,
  author  = {Edwin C. Y. Koh},
  title   = {From Text to {DSM}: Evaluating the Impact of Writing Style and Entity Naming on {LLM}-Based Retrieval of Asymmetrical Indirect Design Dependencies},
  journal = {Research in Engineering Design},
  volume  = {37},
  pages   = {13},
  year    = {2026},
  doi     = {10.1007/s00163-026-00476-2}
}

@article{Koh2024,
  author  = {Edwin C. Y. Koh},
  title   = {Auto-{DSM}: Using a Large Language Model to Generate a Design Structure Matrix},
  journal = {Natural Language Processing Journal},
  volume  = {9},
  pages   = {100103},
  year    = {2024},
  doi     = {10.1016/j.nlp.2024.100103}
}

@inproceedings{Langner2025,
  author    = {Christoph Langner and Yevheniya Paliyenko and Dominik Roth and Matthias Kreimeyer},
  title     = {Utilizing {DSM} and {SysML} for Modeling Data Flows in Complex Networks -- A Case Study on Autonomous Public Transportation},
  booktitle = {Proceedings of the 27th International DSM Conference (DSM 2025), DS 141},
  pages     = {11--20},
  year      = {2025},
  doi       = {10.35199/dsm2025.02}
}

@article{LeichtNewman2008,
  author  = {E. A. Leicht and M. E. J. Newman},
  title   = {Community Structure in Directed Networks},
  journal = {Physical Review Letters},
  volume  = {100},
  pages   = {118703},
  year    = {2008},
  doi     = {10.1103/PhysRevLett.100.118703}
}

@article{Luo2015,
  author  = {Jianxi Luo},
  title   = {A simulation-based method to evaluate the impact of product architecture on product evolvability},
  journal = {Research in Engineering Design},
  volume  = {26},
  pages   = {355--371},
  year    = {2015},
  doi     = {10.1007/s00163-015-0202-3}
}

@misc{OpenAI2025,
  author       = {OpenAI},
  title        = {{GPT-5.2} System Card},
  year         = {2025},
  howpublished = {\url{https://openai.com/index/gpt-5-system-card-update-gpt-5-2/}}
}

@inproceedings{Pimmler1994,
  author    = {Thomas U. Pimmler and Steven D. Eppinger},
  title     = {Integration Analysis of Product Decompositions},
  booktitle = {Proceedings of the ASME Design Theory and Methodology Conference},
  pages     = {343--351},
  year      = {1994},
  doi       = {10.1115/DETC1994-0034}
}

@misc{QwenTeam2026,
  author       = {{Qwen Team}},
  title        = {{Qwen3.5}: Towards Native Multimodal Agents},
  year         = {2026},
  howpublished = {\url{https://qwen.ai/blog?id=qwen3.5}}
}

@inproceedings{Roh2025,
  author    = {Hongman Roh and Lena Etzenbach and Alexandre Oltramare and Jonas Norheim and Olivier de Weck},
  title     = {Factored Dependency Structure Matrix for Representation of Multi-Connection Systems},
  booktitle = {Proceedings of the 27th International DSM Conference (DSM 2025), DS 141},
  pages     = {31--40},
  year      = {2025},
  doi       = {10.35199/dsm2025.04}
}

@article{RomeraParedes2024,
  author  = {Bernardino Romera-Paredes and Mohammadamin Barekatain and Alexander Novikov and Matej Balog and M. Pawan Kumar and Emilien Dupont and Francisco J. R. Ruiz and Jordan S. Ellenberg and Pengming Wang and Omar Fawzi and Pushmeet Kohli and Alhussein Fawzi},
  title   = {Mathematical Discoveries from Program Search with Large Language Models},
  journal = {Nature},
  volume  = {625},
  pages   = {468--475},
  year    = {2024},
  doi     = {10.1038/s41586-023-06924-6}
}

@inproceedings{Solberg2025,
  author    = {Ragnar Solberg and Ali Yassine and Nicolay Worren and Kjetil Soldal and Thomas Christiansen},
  title     = {{DSM}s for Organization Design: Incorporating Additional Criteria in Clustering Algorithms},
  booktitle = {Proceedings of the 27th International DSM Conference (DSM 2025), DS 141},
  pages     = {89--98},
  year      = {2025},
  doi       = {10.35199/dsm2025.10}
}

@article{Sosa2004,
  author  = {Manuel E. Sosa and Steven D. Eppinger and Craig M. Rowles},
  title   = {The Misalignment of Product Architecture and Organizational Structure in Complex Product Development},
  journal = {Management Science},
  volume  = {50},
  pages   = {1674--1689},
  year    = {2004},
  doi     = {10.1287/mnsc.1040.0289}
}

@article{Steward1981,
  author  = {Donald V. Steward},
  title   = {The Design Structure System: A Method for Managing the Design of Complex Systems},
  journal = {IEEE Transactions on Engineering Management},
  volume  = {EM-28},
  pages   = {71--74},
  year    = {1981},
  doi     = {10.1109/TEM.1981.6448589}
}

@mastersthesis{Thebeau2001,
  author = {Russell E. Thebeau},
  title  = {Knowledge Management of System Interfaces and Interactions for Product Development Processes},
  school = {Massachusetts Institute of Technology},
  year   = {2001}
}

@inproceedings{Yang2024,
  author    = {Chengrun Yang and Xuezhi Wang and Yifeng Lu and Hanxiao Liu and Quoc V. Le and Denny Zhou and Xinyun Chen},
  title     = {Large Language Models as Optimizers},
  booktitle = {International Conference on Learning Representations (ICLR 2024)},
  year      = {2024}
}

@article{Yu2007,
  author  = {Tian-Li Yu and Ali A. Yassine and David E. Goldberg},
  title   = {An Information Theoretic Method for Developing Modular Architectures Using Genetic Algorithms},
  journal = {Research in Engineering Design},
  volume  = {18},
  pages   = {91--109},
  year    = {2007},
  doi     = {10.1007/s00163-007-0030-1}
}

\appendix
\section{Implementation Details}
\label{app:implementation}

\subsection{Prompt Details}
\label{app:prompt}

Each LLM query consists of a system message and a user message. The
system message establishes the task context: the LLM is instructed to
act as a DSM modularization expert whose goal is to find a partition of
system elements that minimizes the structural cost objective. The user
message contains three structured blocks.

\textbf{Block 1: DSM description.} Under $k=0$, nodes are labeled with
anonymized identifiers (e.g., N01, N02, \ldots) in randomized order.
Under $k=1$, each node is presented with its engineering name and a
one-sentence functional description (e.g., `\texttt{N01 (Fuel System):
Stores and delivers fuel to the propulsion subsystem}'). The DSM
structure is given as a directed edge list: `\texttt{N03~-{}-{}> N01
(weight: 4)}', one edge per line. Edges are shuffled randomly at each
iteration to prevent the LLM from relying on list position as a proxy
for node importance.

\textbf{Block 2: Solution base.} Up to 10 previously evaluated
partitions are presented, sorted in ascending order of TotalCost (best
first). Each solution is shown as a mapping from node to module label
(e.g., `\texttt{N01: M2, N02: M1, N03: M2, \ldots}') followed by its
TotalCost value. In the baseline configuration, the pool contains the 5
best solutions found so far and 5 randomly sampled solutions from the
full history. In the objective formulation ablation
(Section~\ref{subsec:ablation-design}), a formula variant additionally
states the TotalCost formula explicitly; in the no-formula variant, only
the ranked examples are provided.

\textbf{Block 3: Generation instruction.} The LLM is instructed to
generate a new partition that achieves lower TotalCost than the best
solution shown, to output the result in a parseable format (JSON
dictionary mapping node label to module label), and not to copy any of
the provided solutions verbatim.

\subsection{Parsing and Validation}

The LLM response is parsed by extracting the first JSON-like block from
the output using a regular expression. The extracted mapping is validated
against three criteria: (1)~all $n$ node identifiers are present and
assigned to exactly one module; (2)~at least 2 distinct module labels
are used; and (3)~all values are string labels rather than numerical indices. If any criterion fails,
the response is discarded and the iteration is recorded as invalid.
Module labels are then canonicalized to consecutive integers
($M_1, M_2, \ldots, M_K$) before TotalCost evaluation.

\subsection{SA Reference Computation}
\label{app:sa}

The SA reference for each case is computed using simulated annealing
with a geometric cooling schedule ($\alpha=0.9$, 150 temperature steps).
Each run starts from a random initial partition; at each step, a
randomly selected element is proposed for reassignment to an existing or
new module, with the move accepted probabilistically according to the SA
acceptance criterion. A total of 10{,}000 independent random restarts
are executed, and the SA reference is the minimum TotalCost observed
across all restarts. SA reference values are reported in
Table~\ref{tab:results}.

\subsection{Hyperparameter Settings}
\label{app:hyperparams}

Table~\ref{tab:hyperparams} summarizes the key hyperparameters used in
all experiments. The values listed correspond to the default configuration for the
main experiments (Table~\ref{tab:results}) and are held constant across
all models, cases, and knowledge conditions unless explicitly varied as
part of the ablation study (Section~\ref{subsec:ablation-design}).

\begin{table}[H]
\caption{LLM-CO hyperparameter settings}
\label{tab:hyperparams}
\centering
\begin{tabular}{lc}
\toprule
\textbf{Hyperparameter} & \textbf{Value} \\
\midrule
Number of iterations              & 30 \\
Number of independent runs        & 10 per condition \\
Solution pool: best kept ($p$)    & 5 \\
Solution pool: random added ($q$) & 5 \\
TotalCost exponent ($\rho$)        & 1 \\
LLM temperature                   & 1.0 \\
Default input format              & Directed edge list \\
Default knowledge condition       & $k=0$ and $k=1$ (both reported) \\
\bottomrule
\end{tabular}
\end{table}

\end{document}